\documentclass[useAMS,usenatbib,usegraphicx]{mn2e}
\usepackage{amssymb}
\usepackage{amsmath}
\usepackage{times}

\newcommand{\kms}{\mbox{\,km s$^{-1}$}}
\newcommand{\pasp}{PASP}
\newcommand{\apjs}{ApJS}
\newcommand{\apj}{ApJ}

\newcommand{\aap}{A\&A}
\newcommand{\aj}{AJ}
\newcommand{\mnras}{MNRAS}

\title[New Local Group galaxy VV124]{VV124 (UGC4879): A new transitional dwarf galaxy in the periphery of the Local Group}

\author[Kopylov et al.]{A. I. Kopylov$^1$\thanks{E-mail: akop, ntik, fabrika, azamat@sao.ru (AIK, NAT, SF, AFV); dio@iac.es (ID)}
N. A. Tikhonov$^1$, S. Fabrika$^1$, I. Drozdovsky$^{2,3}$, and A. F. Valeev$^1$ 
\\
$^1$Special Astrophysical Observatory, Nizhnij Arkhyz 369167, Russia\\
$^2$Instituto de Astrof\'{\i}sica de Canarias, C/V\'{\i}a Lactea s/n,38200, La Laguna, Tenerife, Spain\\
$^3$Astronomical Institute of St.Petersburg State University, Russia}

\begin{document}

\date{\today}
\pagerange{\pageref{firstpage}--\pageref{lastpage}} \pubyear{2008}
\maketitle
\label{firstpage}

\begin{abstract}
We present the first resolved-star photometry of VV124 (UGC4879) and 
find that this is the most isolated dwarf galaxy in the periphery of 
the Local Group. 
Based on imaging and spectroscopic follow up observations with the 6m BTA 
telescope, we resolve VV124 into 1560 stars down to the limiting magnitude 
levels of $V\simeq25.6$ and $I\simeq23.9$. 
The young blue stellar populations and emission gas are found near the core, 
but noticeably displaced from the center of the galaxy as traced by dominant 
evolved red stars. The mean radial velocity derived from the spectra of two 
Blue Supergiant stars, an H{\sc ii} region and unresolved continuum sources 
is $-80\pm10$\,km/s. The evolved ``red tangle'' stellar 
populations, which contains the red giant branch (RGB), are identified 
at large galactocentric radii. 
We use the $I$-band luminosity function to determine the distance based on 
the Tip of RGB method, $1.1\pm 0.1$\,Mpc. This is $\sim10$ times closer than 
the values usually assumed in the literature, and we provide revised distance 
dependent parameters. From the mean $(V-I)$ color of the RGB, we estimate 
the mean metallicity as [Fe/H]$\simeq-1.37$ dex.  
Despite of its isolated location, 
the properties of VV124 are clearly not those of a galaxy in formation, 
but rather similar to a transitional dIrr/dSph type.
\end{abstract}
\maketitle

\section{Introduction}
\label{sec:intro}


The star formation and chemical enrichment histories (SFH) of the  Local
Group galaxies, which can be derived from their resolved 
stellar populations, directly test cosmological galaxy formation models. 
However, the dynamical and mass loss histories of the nearest
satellite galaxies are the major uncertainty in understanding their
evolution \citep[e.g., ][]{Mayer+06}. Isolated dwarf galaxies are not
affected by these environmental effects, and are therefore ideal probes
of the basic mechanisms affecting the SFH of a galaxy. Galaxies in the
periphery of the Local Group (LG) are particularly valuable in this
respect, being the only {\em isolated} systems for which detailed
information is possible on their complete SFH with the 
Hubble Space Telescope 
\citep[HST; ][]{Gallart+07}. Additionally, these fringe dwarfs are important
probes of the dynamical state of the local universe\citep{vdMarel+07}. 
The purpose of this Letter is to announce the discovery of an additional 
exceptionally isolated galaxy in the outskirts of the LG. 

One of the authors (AIK) has noticed a discrepancy between the distance 
based on adopted radial velocity 
of VV124 (UGC4879), $V_h=600\kms$, 
cited in both the NED and HyperLeda databases, and the galaxy's apparent 
resolution into
stars on the Sloan DSS images. The bright blue stars are well distinguished
near the center of VV124, while a few dozen fainter red objects can be traced 
to large galactocentric distances.  
Assuming this field population is composed
of red giant (RGB) and asymptotic giant branch (AGB) stars,
we made a preliminary estimation of VV124 distance as $D\leq 2$\,Mpc.

This object is one of the Vorontsov-Veliaminov galaxies listed in his ``Atlas
and Catalog of Interacting galaxies'' \citep{Vorontsov} as a ``Nest'' type.
Later the galaxy was included in \citeauthor{Zwicky:68}'s (1968; UZC) 
and \citeauthor{Nilson:73}'s galaxy catalogs (1973; UGC). 
The later catalog noted a presence 
of ``several faint very blue condensations superimposed'' in the object.
\citet{Jansen1} described VV124 as a low-luminosity dwarf that ``shows the 
enhanced Balmer absorption lines and blue continuum of a young 'poststarburst' 
galaxy.'' 

The radial velocity measurement of $V_h=600\pm100$\kms for VV124 comes from  
the first CfA redshift catalog \citep{Huchra}, that was reproduced in the RC3 
catalog with uncertainty of $50$\kms, and currently cited in both NED and
HyperLeda databases with a note on alternative measurement in UZC \citep{Falco},
$V_h=62\pm69$\kms. The revised 2002 version of UZC gives $V_h=-44\pm69$\kms, 
derived via cross-correlation of the spectra obtained
with the help of the FAST spectrograph on the 1.5m Tillinghast telescope.      
This measurement however remained unnoticed, and the galaxy was even considered 
by \citet{Azzaro+06} as a distant companion of NGC2841, 
which has a Cepheid distance of $\sim14$\,Mpc \citep{Macri+01,Saha+06}. 
VV124 is undetected in HI \citep{Schneider}, probably due to the survey's
search range, $100-6800$\kms.  
Another possible reason behind the lack of concerns over the VV124 distance, is 
a relatively high surface brightness, $\mu_{B_{25}}=23.55$,
for its new distance-corrected luminosity, $M_B=-11.6$ --about 1.5~mag above 
the value expected from the empirical absolute magnitude versus surface 
brightness relationship \citep[e.g, ][]{Kar}. 
Various optical \citep{Jansen2,Taylor}
and near-IR imaging surveys \citep{Grauer} aimed at the global photometric 
parameters and also missed an apparent high degree of VV124 resolution.

\begin{figure}
\includegraphics[width=\columnwidth,angle=0,trim=0mm 0mm 0mm 0mm]{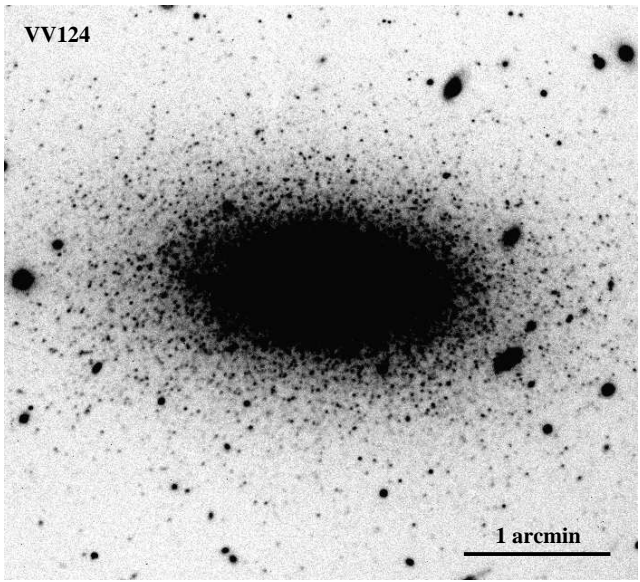}
\caption{
Image covering a $4.3\arcmin \times 4.3\arcmin$ region centered on VV124
made from the 6m telescope data in V band. North is up, and 
east is to the left.
}
\label{1}
\end{figure}


\begin{figure}
\includegraphics[width=\columnwidth,angle=0,trim=0mm 0mm 0mm 0mm]{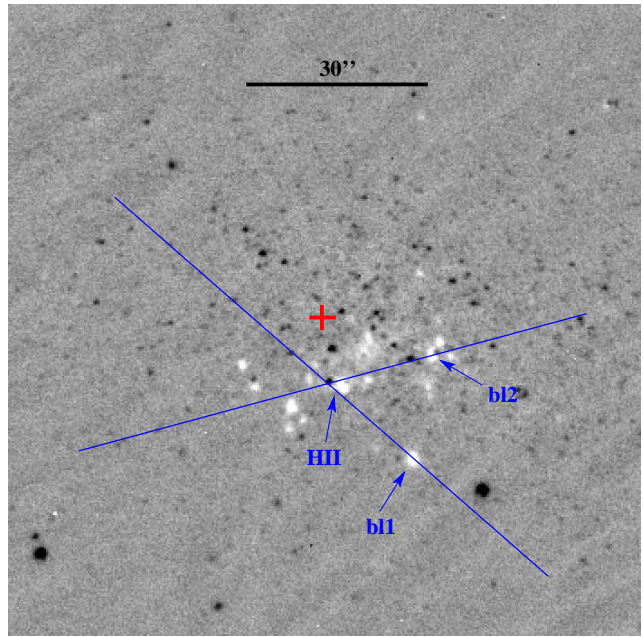}
\caption{
The distribution of blue (in white) and red stars (in black) in the central 
$1.5\arcmin \times 1.5\arcmin$ area, obtained by division of I-to-V band 
images. The lines mark the slit positions for the spectroscopic 
observations. The red plus sign shows the center of VV124.  
Two analyzed Blue Supergiant stars, and an H{\sc ii} region 
are indicated.}
\label{2}
\end{figure}

\section{Observations and Stellar Photometry}

Follow-up imaging observations of VV124 were obtained on 2008 January 10/11 
using the SCORPIO \citep[Spectral Camera with Optical Reducer for Photometrical and 
Interferometrical Observations; ][]{Afanasiev} mounted at the prime focus of 
the 6m BTA telescope of the Special Astrophysical Observatory (Russia). 
The detector
was a EEV42-40 $2048\times2048$ CCD array with a field of view roughly 
$6\times6$ arcmin and a scale of $0.357\arcsec$ per $2\times 2$-binned 
pixel. VV124 was observed with $3\times200$\&$3\times60$ s in $V$ and 
$2\times200$\&$3\times60$ in $I$ broad-band filters close to
the standard John\-son-\-Cou\-sins photometric system. The weather conditions 
were good, with a stable seeing of $1.0-1.1\arcsec$. 
The primary data reduction was performed using the ESO-MIDAS software 
package. It included de-biasing, flat fielding and cosmic-ray hit removal. 

\begin{figure}
\includegraphics[width=\columnwidth,angle=0]{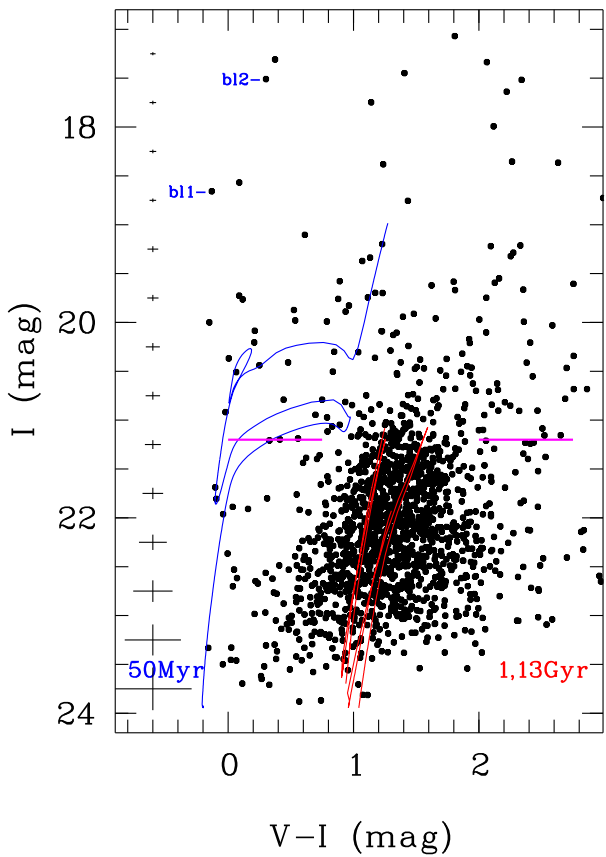}
\caption{
Color-magnitude diagram, $V-I$ vs. $I$, of the VV124 resolved stars, based
on 6-m BTA imaging data. The error bars show the binned mean standard errors
of the photometry.
The Tip of the first ascent Red Giant Branch (TRGB)
is marked with 2 horizontal lines. The isochrones are for the 
metallicity $0.001\,Z\odot$. 
}
\label{3}
\end{figure}

Figure~1 shows a coadded image from these data, exhibiting many resolved 
stars. The image of the central part of the galaxy obtained
by division of $I$-on-$V$-band frames  
is shown on Fig.~2, and it demonstrates a slight asymmetry in the distribution
of the blue stars (white on the figure) relative to the red stars, which
can be traced to large galactocentric distances. 


We performed single-star photometry with DAOPHOT~II \citep{Stetson}.
Because of relatively strong image aberrations of the SCORPIO camera, 
we carefully selected isolated stars suitable to model the point-spread 
function (PSF), and removed stars from the detection catalog located 
in the areas where the PSF reconstruction was inadequate.
The absolute photometric calibration of the data was performed
using the observations of standard stars from the
lists by \citet{Landolt} on the same nights.
We corrected data for foreground
extinction in the direction of VV124 using the values of 
\citet{Schlegel}: $A_V = 0.050$ and $A_I = 0.029$. 
We did not correct for
internal extinction within VV124, but the transparency of the outer field
stellar population (the population from which we derive the distance) is high 
and suggests that the extinction is low at large galactocentric radii.

The color-magnitude diagram (CMD) of VV124 (Fig.~3) illustrates a stellar 
mix that is characteristic of the superposition of many stellar generations,
and reminiscent of the CMD of a typical transitional --intermediate  
dIrr/dSph galaxy like Phoenix or LGS3 \citep{Gallart+04, Gallart+07}. 
The ``blue plume" 
at $(V-I)\sim0$ contains massive blue supergiants and main-sequence stars, 
the ``red plume" 
at $(V-I)\sim1.5$ contains evolved giants
and asymptotic giant branch (AGB) stars, while the ``red tail" extending 
past $(V-I)>2$ contains intermediate-mass AGB stars in the thermally pulsing 
phase, and the concentration of stars below the TRGB at $I \simeq 21.2$ 
contains low-mass red giants (RGB) and AGB stars. For reference, we overlay 
the CMD with the Z=0.001 BaSTI stellar 
isochrones \citep{Pietrinferni+04} assuming the distance derived below. 
The blue plume is quite scarce, whereas the RGB exhibits a significant width.
The stellar crowding and blending, as well as a fluctuation of 
internal extinction, may dominate these effects. However based on
preliminary artificial star tests, the RGB width
exceeds that photometric errors, suggesting a possible wide range of stellar 
metallicities and/or ages.  

\section{Distance and metallicity}
We used our stellar photometry data to find the TRGB and to determine
the distance and metallicity of VV124 (following \citet{Lee}).
The photometric data of outer stars are preferable over the central high 
stellar density region since they exhibit much less crowding and fewer
AGB stars, which tend to have a stronger concentration towards the 
center of irregular galaxies \citep{Tikhonov1,Tikhonov2}.
Therefore, for the analysis of the RGB we exclude the innermost region 
defined by a central circle with diameter of $90\arcsec$. 
There is a sufficiently large number of stars detected in the selected area 
to have $0.7<(V-I)_0<2.0$. The TRGB was found by applying a Sobel filter
to the $I$-band luminosity function; 
it occurs at $I_0 = 21.17\pm0.10$ (indicated in Fig.~3 by a line). 
In order to estimate the average metallicity of the RGB stars we measure 
the color 
at half a magnitude below the TRGB, $(V-I)^0_{-3.5} = 1.43$, corresponding
to the $(V-I)^0_{TRGB} = 1.63$. 
Based on the calibration by \citet{Lee} we estimated the 
metallicity of RGB stars of [Fe/H]$ = -1.37$, and a distance of 
$D = 1.1\pm 0.1$\,Mpc. 
 
Note that the result on the metallicity is based on the 
assumption that the age of the RGB stars is predominantly old, since
the calibration of \citet{Lee} is based on old galactic globular clusters,
therefore, if the majority of stars are younger than $10-13$\,Gyr, this 
metallicity estimation is an upper limit of an intrinsic value.  

\section{Spectral Observations}

Long-slit spectral observations of VV124 were obtained with the 
same 6m/SCORPIO facility on 2008 February 6/7 in a wavelength range 
of $4000-5600$\,\AA\/ with
spectral resolution of $5$\,\AA. Two $1\arcsec$-width long slits were
placed on two bright blue stars near the center of the galaxy, marked on
Fig.~2 as 'bl1' and 'bl2'. 
With the VV124 distance of 1.1~Mpc, the scale along the 
slit corresponds to 5.3~pc/$\arcsec$. This gives a separation of 111\,pc
between stars 'bl1' and 'bl2'. 
Both slits run over the compact H{\sc ii}-region, while the 
second one also includes by chance a high-redshift background  
galaxy.
 
The spectral data have been reduced using the standard calibration procedures
within the IDL environment. The uncertainty of the wavelength calibration 
is 10\kms. However, the final accuracy of the radial velocity measurements 
was limited by the quality of the spectrum of each of the detected components. 

By matching the spectra of the two bright stars with the template
spectra from the STELIB stellar library 
\citep[http://www.ast.obs-mip.fr/article181.html]{LeBorgne+03}, we
find that both stars are Blue Supergiants (BSG) of spectral 
type O9.5 for 'bl1', and F5Ia ('bl2'). 
HD30614 ($\alpha$\,Cam, O9.5Iae) and HD188209 (O9.5Ib) are good matches 
for the star 'bl1', and the spectrum of HD269697 (F5Ia) from the Large 
Magellanic Cloud is a close match of that of 'bl2'.
From the results of our stellar photometry we find that 'bl1' has 
$V = 18.53$ and $(V-I)= -0.13$, while magnitude and color of 'bl2' are
$V = 17.81$ and $(V-I)= +0.30$. 
This corresponds to absolute magnitudes
of $M_V = -6.73$ and $M_V = -7.45$ for 'bl1' and 'bl2', respectively.

To measure the radial velocities, we used absorption lines of hydrogen, 
H$\delta$, H$\gamma$ and H$\beta$,
because these are the deepest and most reliable lines for redshift
measurements. The radial velocity of 'bl1' is $-79 \pm 10$\kms, while 
radial velocity of 'bl2'  is $-76 \pm 15$\kms.

In addition, we analyzed the redshifts of absorption hydrogen lines of  
the unresolved stellar populations integrated at various areas 
along the slits. These hydrogen lines are the most prominent 
details in the spectra of unresolved objects.  
The mean value of the radial velocity obtained from these spectra
$-86 \pm 20$\kms, is in good agreement with the measured 
velocities of the two BSGs. 

We also found a compact H{\sc ii} region exhibiting faint H$\beta$ emission
(see Fig.~2), as well as 
an extended region of faint diffuse [OIII]$\lambda 5007, 4959$ 
emission, located in the north-east part of the galaxy, traced over 70\,pc
along the first slit and over 120 pc in the second slit.  
We will present a more detailed analysis of the spectroscopic 
data in a follow-up work. 

\section{Results and discussion}



\begin{table}
\caption{Properties of the VV124 dwarf}
\label{data}
\bigskip
\begin{tabular}{|l|c|l|} \hline
Parameter          & Value            & Reference       \\
\hline
\hline
 $A_B$          & 0.07             & NED           \\
 $B_T$          & $13.68\pm0.03$ & \citet{Taylor}   \\
 $K_T$          & $12.05       $ & \citet{Grauer}   \\
Size$_{25}$     & $2\arcmin.1\times1.3$ & HyperLeda \\
 P.A.           & $87\degr.5$      & Leda      \\
 $(U-B)_T$      & $-0.08\pm0.04$ & \citet{Taylor} \\
 $(B-V)_T$      & $ 0.58\pm0.04$ & \citet{Taylor} \\
 $(B-R)_T$      & $ 0.85\pm0.05$ & \citet{Jansen1} \\
\hline
Distance (Mpc)  & $1.1\pm0.1$      & this work \\
$ [Fe/H]$         & $-1.37$          & this work \\
 $V_h$ (\kms)   & $-80 \pm 10$     & this work \\
$V_{LG}$ (\kms) & $-22$            & this work \\
 $M_B$          & $-11.6$          & this work \\
 $M_K$          & $-13.2$          & this work \\
Size$_{25}$ (kpc) & $0.67\times0.41$ & this work \\
\hline
\end{tabular}
\end{table}

The basic parameters of VV124 are presented in Table~\ref{data}, based
on our measurements and data available from the literature. 
As judged from our distance estimation and new photometric and spectroscopic 
measurements, VV124 lies at a distance of 1.1~Mpc and appears
to be a transitional type dwarf galaxy: although possessing some young 
stellar populations the bulk of stars is evolved.
Despite its isolated location, and some recent star formation, the properties 
of VV124 are clearly not those of a galaxy in formation.  

VV124 is located  $1.1$\,Mpc from the MW and $1.2$\,Mpc from M31 --on 
the periphery of the Local Group (Fig.~4). Leo~A is its nearest neighbor, 
at a separation of $\simeq0.5$~Mpc. There are only two other
galaxies in the LG with similar, although slightly lower, degree of 
isolation: DDO~210 and SgrDIG, located in the direction opposite 
to that of VV124. 
Remarkably,
the peculiar radial velocity of these three galaxies, 
corrected either for the LG rest frame or for the motion within 
the Local Sheet --following relation (15) from \citet{Tully+08}-- 
is close to zero, $-22<V_{LG}<+10$ and $-5<V_{LSh}<+9$\kms, 
confirming that these objects are near the LG turn-around radius.
Most importantly, these galaxies are not, and have never been, 
satellites of either of the dominant members of the Local Group. Their
free-fall time into M31 or the MW, 
is longer than a Hubble time. The exceptional isolation means that
these are among the few galaxies in the nearby Universe which evolution
have never been complicated by the local environmental mechanisms.
Therefore, these objects are ideal probes of the basic mechanisms 
affecting the star-formation history of a galaxy. This would, however,
require deeper observations, preferably with HST. 
 


\begin{figure}
\includegraphics[width=\columnwidth,angle=0]{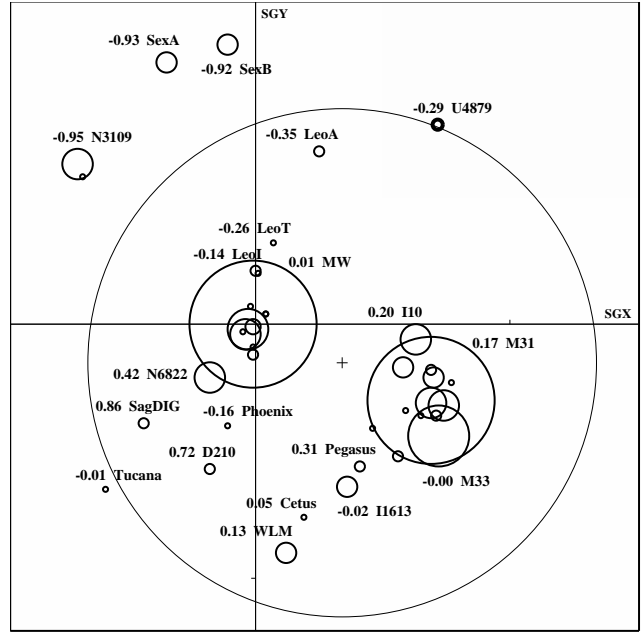}
\caption{
Updated map of the Local Group and its surroundings in
Supergalactic SGX, SGY coordinates. 
The numbers show the SGZ-distance in Mpc to the SGXY-plane. The large 
circle of 1\,Mpc radius is
centered midpoint between the Milky Way and Andromeda, which are
indicated with two 0.25\,Mpc radius circles. 
VV124 (UGC4879) is visible in the upper right.
}
\label{4}
\end{figure}

\section*{Acknowledgments}

Based on observations obtained with 6m BTA telescope (Special
Astrophysical Observatory of Russian Academy of Sciences). 
We acknowledge the usage of Sloan Digital Sky Survey, and
the HyperLeda, NED and SIMBAD databases. 
ID acknowledges the support from the IAC program P3/94 and the 
Spanish Ministry of Science and Technology grant AYA 67913.




\label{lastpage}
\end{document}